\documentclass[preprintnumbers, amsmath, amssymb, showpacs, showkeys, twocolumn, superscriptaddress, 
nofootinbib]{revtex4}

\usepackage{graphicx,subfigure,bm,color,psfrag,hyperref}

\usepackage{mathtools}

\usepackage{amsmath,amsthm}
\usepackage{amsfonts}
\usepackage{amssymb}
\usepackage{verbatim}
\usepackage{mathrsfs}
\usepackage{tensor}
\usepackage[dvipsnames]{xcolor}
\usepackage{newpxmath}
\usepackage[scaled=.95]{newpxtext}
\usepackage[all]{xy}

\newcommand{\lin}{\\[7pt]}

\newcommand{\pder}[2]{\dfrac{\partial#1}{\partial#2}}
\newcommand{\pdder}[3]{\dfrac{\partial^2 #1}{\partial #2 \partial #3}}

\newcommand{\dder}[2]{\dfrac{\delta#1}{\delta#2}}
\newcommand{\pdot}[1]{\dot{\partial}_{#1}}

\newcommand{\Gd}{\mathcal{G}}

\newcommand{\R}{\mathcal{R}}
\newcommand{\de}{\mathrm{d}}

\begin{document}

\title{Bounce cosmology in generalized modified gravities}

 \author{G. Minas}
 \email{geminas@phys.uoa.gr}
 \affiliation{Section of Astrophysics, Astronomy and Mechanics, Department of Physics,  
National and Kapodistrian University of Athens, Panepistimiopolis 15783,
Athens, Greece}

 \author{E. N. Saridakis}
\email{msaridak@phys.uoa.gr}
\affiliation{Department of Physics, National Technical University of Athens, Zografou
Campus GR 157 73, Athens, Greece}
\affiliation{Department of Astronomy, School of Physical Sciences, University of Science 
and Technology of China, Hefei 230026, China}

 \author{P. C. Stavrinos}
 \email{pstavrin@math.uoa.gr}
 \affiliation{  Department of Mathematics, National and Kapodistrian University of Athens,
Panepistimiopolis 15784, Athens, Greece}
 
 \author{A. Triantafyllopoulos}
 \email{alktrian@phys.uoa.gr}
 \affiliation{Section of Astrophysics, Astronomy and Mechanics, Department of Physics,  
National and Kapodistrian University of Athens, Panepistimiopolis 15783,
Athens, Greece}

\begin{abstract}
We investigate the bounce realization in the framework of generalized modified 
gravities arising from Finsler and Finsler-like geometries. In particular,  the richer 
intrinsic geometrical
structure is reflected in the appearance of extra degrees of freedom in the Friedmann 
equations that can drive the bounce. We examine various Finsler and Finsler-like 
constructions. In the cases of general very special relativity as well as of Finsler-like 
gravity on the tangent bundle we show that a bounce cannot be easily obtained. However, 
in the Finsler-Randers space the induced scalar anisotropy can fulfill  the bounce 
conditions and bouncing solutions are easily obtained. Finally, for the general 
class of theories that include a nonlinear connection a new scalar field is induced,
leading to a scalar-tensor structure that
can easily drive a bounce. These features reveal the capabilities of Finsler and 
Finsler-like geometries.
\end{abstract}


\maketitle

\section{Introduction}

Bounce cosmologies offer an alternative view of the early universe
\cite{Biswas:2005qr,Cai:2007qw,Cai:2008qw,Cai:2012va,Cai:2014bea,Brandenberger:2016vhg} 
(for a review see \cite{Novello:2008ra}). Historically, this idea belongs to Tolman who  
in 1930's first suggested the possibility of a re-expansion of a closed universe which 
has 
already collapsed to an extremely dense state \cite{Tolman:1931fei}.
Since then, various bouncing models have been proposed within an effort for a systematic 
explanation of the origin of our universe. 

The main advantage of bouncing cosmology is that it provides a way of solving the 
singularity problem which appears in the standard cosmological paradigm. The singularity 
(Big 
Bang) is replaced with a smooth transition from contraction to expansion (Big Bounce). In 
this sense, bounce cosmology offers an opportunity of obtaining a more continuous picture 
of the early universe. The efficiency of bouncing models in solving basic cosmological 
problems in comparison with inflationary scenarios is visualized via the wedge diagram 
introduced in \cite{Ijjas:2018qbo}.

In general, the realization of a bounce requires a violation of the null energy 
condition. This can be achieved with the introduction of  extra degrees of freedom which 
are added ad hoc into the Lagrangian \cite{Singh:2015bca,Cai:2012va}. The violation of 
null energy condition needs to be handled with care, in order not to spoil the 
usual thermal history and the sequence of epochs after the bounce. Nevertheless, such 
violations can easily be acquired from modified \cite{Novello:2008ra} or quantum gravity 
\cite{Barrau:2016fcg}. In particular, they can be easily acquired  in the Pre-Big-Bang 
\cite{Gasperini:2003pb,Gasperini:2004ss} and the Ekpyrotic \cite{Khoury:2001wf,Khoury:2001bz} models, in
gravity actions with higher 
order corrections \cite{Tirtho1,Nojiri:2013ru}, in $f(R)$ gravity 
\cite{Bamba:2013fha,Nojiri:2014zqa}, in $f(T)$ 
gravity 
\cite{Cai:2011tc}, in braneworld scenarios \cite{Shtanov:2002mb,Saridakis:2007cf}, in
non-relativistic gravity \cite{Brandenberger:2009yt,Cai:2009in,Saridakis:2009bv}, in 
Galileon theory \cite{Easson:2011zy,Qiu:2013eoa}, in 
massive gravity  
\cite{Cai:2012ag},  in Lagrange 
modified gravity \cite{Cai:2010zma}, in loop quantum 
cosmology \cite{Bojowald:2001xe,Odintsov:2014gea,Odintsov:2015uca}, etc. Moreover, a 
nonsingular bounce model which supports magnetogenesis at the inflationary epoch is
presented in \cite{Membiela:2013cea}.

Among modified gravity theories, an interesting class is that of gravitational models 
based on Finsler and Finsler-like geometries. These are natural extensions of Riemannian 
geometry in which the physical quantities may directly depend on observer 4-velocity, and 
this velocity-dependence reflects the Lorentz-violating character of the kinematics. Such 
a property is called \emph{dynamic anisotropy} 
\cite{Bogoslovsky:1999pp,Chang:2007vq,Kouretsis:2010vs,Vacaru:2010fi,Mavromatos:2010jt,
Vacaru:2010rd,Mavromatos:2010nk,Torrome:2012kt,Basilakos:2013hua,Basilakos:2013ij, 
Hohmann:2016pyt,Hohmann:2018rpp}. 
Additionally, Finsler and Finsler-like geometries are strongly connected to the effective 
geometry within anisotropic media \cite{Born,Perlick} and naturally enter the analogue 
gravity program \cite{Barcelo:2005fc}. These features suggest that Finsler and 
Finsler-like geometries may play an important role within quantum gravity physics. The 
dependence of the metric tensor and other quantities on the position coordinates of the 
base-manifold and the directional/velocity variables of the tangent space suggest that 
the 
natural geometrical framework for the description of these models is the tangent bundle 
of 
a smooth manifold. Finally, in the case where there is no velocity-dependence, Finsler 
geometry becomes Riemannian.
 
The intrinsic geometrical space-time dynamical anisotropy of Finsler geometry (not to be 
confused 
with the spatial anisotropy that may exist also in Riemannian geometry, as for instance 
in Bianchi cases) is included in the geometry of space-time as an intrinsic field 
(variable) which influences its   geometrical and physical concepts. Hence, it 
can give us the form of anisotropy as a hypothetical field, the \emph{anisotropion}, 
which produces this deviation from isotropy. This appears in the Friedmann equations 
and Lorentz violations   \cite{Koivisto:2008ig,Stavrinos:2012ty,kour-stath-st 
2012,Triantafyllopoulos:2018bli}, and thus the anisotropy arises as a property of 
Finslerian spacetime
\cite{stavrinos-ikeda 1999,Stavrinos:2016xyg,Stavrinos:2012ty,Triantafyllopoulos:2018bli}.

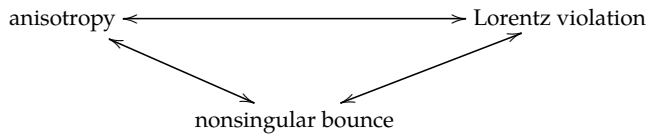
\begin{figure}[h]
\begin{displaymath}
    \xymatrix{ \text{anisotropy} \ar@{<->}[rr] \ar@{<->}[dr]                             
& 
  & \text{Lorentz violation} \ar@{<->}[dl] \\
   & \text{nonsingular bounce} &                                        }
\end{displaymath}
\caption{\it{ Connections of anisotropy, nonsingular bounce and Lorentz 
violation.}}\label{fig:interactions}
\end{figure}

In the present work we are interested in investigating the bounce realization in the 
framework of modified gravity related to Finsler and Finsler-like geometries. In 
particular, we desire to see how the new features of Finsler geometry can drive bouncing 
solutions, and to examine the evolution of the intrinsic anisotropy 
during the bounce. In some bouncing scenarios the anisotropy decreases in the 
contracting 
phase and 
remains quite small during the bounce, in agreement with the current observational data 
\cite{Cai:2012va}. 
On the other hand, there are also scenarios where the reduction of anisotropy in the 
contracting phase is followed by its exponential growth during the 
bounce, mainly due to quantum fluctuations of curvature \cite{Xue:2011nw}. 
Finally, we mention that nonsingular bounces are also possible to be generated in models 
which spontaneously violate 
Lorentz symmetry \cite{kour-stath-st 2012,Gasperini:1985aw,Gasperini:1998eb}. In this framework   Lorentz 
symmetry violations lead to interactions with anisotropies \cite{kour-stath-st 2010}.
Hence, we can establish a connection   between 
anisotropic fields and nonsingular bounce. In summary, we can depict 
all the above form of 
connections in the diagram of Fig. \ref{fig:interactions}.

The plan of this work is the following: In Section \ref{Finslergravity} we first 
describe the basic conditions for a bounce realization and we briefly review Finsler 
geometry and gravity. Then we examine the bounce realization in general very special 
relativity and Finsler-Randers models. In Section \ref{Finslertangentbundle} we study 
the case of Finsler-like gravity on a tangent bundle, while in Section 
\ref{Finslerscalartensor} we analyze bouncing solutions from scalar-tensor 
theory on the fiber bundle. Finally, in Section \ref{Conclusions} we present the summary 
and the conclusions.

\section{Bounce from Finsler gravity}
\label{Finslergravity}

In this section we are interested in studying the bounce realization in the framework of 
Finsler gravity. We start by 
describing the conditions for a bounce realization, and we provide the basics  of 
Finsler geometry and gravity. Then we proceed to the examination of the bounce 
realization in specific models, such as general very special relativity and 
Finsler-Randers ones.

\subsection{Bounce conditions}

Let us start by discussing the basic requirements for a bouncing 
solution. For the moment we consider the ordinary Friedmann-Robertson-Walker (FRW)  
geometry with metric
\begin{equation}
[g_{\mu\nu}(x)] = \mathrm{diag} \left(-1, \frac{a^2(t)}{1-kr^2}, a^2(t)r^2,
a^2(t)r^2\sin^2\theta \right),
\label{FRWdef}
\end{equation}
with $a(t)$ the scale factor 
and $k=-1,0,+1$ corresponding to open, flat and closed spatial geometry respectively. 
As usual, in such a geometry the general field equations of any theory give rise to the 
Friedmann and Raychaudhuri equations, which can be written in a compact form as
\begin{eqnarray}
\label{FR1}
&&H^2=\frac{8\pi G}{3}\rho_{tot}-\frac{k}{a^2}\\
&&\dot{H}=-4 \pi G(\rho_{tot}+P_{tot})+\frac{k}{a^2},
\label{FR2}
\end{eqnarray}
where $G$ is the Newton's constant,  $H={\dot{a}}/{a}$ is the Hubble function, and with 
dots denoting derivatives with 
respect to the cosmic time $t$. In the above expressions $\rho_{tot}$ and $P_{tot}$ are 
respectively the total energy density and pressure of the universe, which include matter, 
radiation, dark energy and any other gravitational or geometrical contribution that a 
theory or scenario may have.

In order to obtain a bounce realization we need a contracting universe, namely with 
$H<0$, succeeded by an expanding universe, namely with $H>0$, and hence from   
continuity   we deduce that at the bounce   point we must have $H=0$. Furthermore, one 
can see that   at the bounce point and around it we must have $\dot H> 0$. Observing 
the form of the 
general Friedmann and Raychaudhuri equations (\ref{FR1}),(\ref{FR2}), and focusing on the 
physically more interesting flat case, we deduce that the 
above requirements can be fulfilled  if  
\begin{eqnarray}
 \rho_{tot}=0
\label{rhototcond1}
\end{eqnarray}
exactly at the bounce point, and if additionally
the null energy condition is violated around 
the bounce point, namely if 
\begin{eqnarray}
 \rho_{tot}+P_{tot}<0
\label{NULLene}
\end{eqnarray}
(in the case of a non-flat universe the bounce can be driven by the curvature term 
without null energy condition violation \cite{Novello:2008ra}). 
Therefore, in order to obtain a bounce one needs to construct theories in which the extra
contributions to the total energy density and pressure are such that the null energy 
condition is violated around the bounce point and requirement (\ref{NULLene}) holds, and 
moreover the total energy becomes zero exactly at the bounce point and condition  
(\ref{rhototcond1}) holds. As we see in the following, scenarios based on Finsler gravity 
can fulfill these necessary conditions.

\subsection{Finsler gravity}

We first  briefly review the basics of Finsler gravity, since 
this lies in the center of the investigation of the present work.
Finsler gravity is a 
geometrical extension of general relativity, where the role of the 
metric is played by the real-valued fundamental function $F(x,y)$, defined on the tangent 
bundle $TM$ over a smooth spacetime manifold $M$. Variable $y$ is an element of the 
tangent space of $M$ at a point $x$ (we have suppressed indices for convenience). The 
distance of two neighboring points on $M$ is defined as $ds=F(x,dx)$. We consider the 
following 
properties to hold:
\begin{enumerate}
 \item $F$ is continuous on $TM$ and smooth on  $ \widetilde{TM}\equiv TM\setminus \{0\} 
$ 
i.e. the tangent bundle minus the null section. \label{finsler field of definition}
 \item $ F $ is positively homogeneous of first degree on its second argument:
  \begin{equation}
   F(x,ky) = kF(x,y), \qquad k>0. \label{finsler homogeneity}
  \end{equation}
 \item The form \begin{equation}f_{\mu\nu}(x,y) = \dfrac{1}{2}\pdder{F^2}{y^\mu}{y^\nu} 
\label{finsler metric} \end{equation} defines a non-degenerate matrix on $TM$ minus the 
null set $ \{(x,y)\in TM | F(x,y)=0\}$: \label{finsler nondegeneracy}
  \begin{equation}
   \det\left[f_{\mu\nu}\right] \neq 0 \label{finsler nondegenerate}.
  \end{equation}
\end{enumerate}
Using homogeneity condition \eqref{finsler homogeneity} it can be shown that:
\begin{equation}
F^2(x,y) = |f_{\mu\nu}(x,y)y^\mu y^\nu|,
\end{equation}
and therefore $f_{\mu\nu}(x,y)$ can play the role of the metric for the vector space 
spanned by $y$. When studying gravity, the metric $f_{\mu\nu}(x,y)$ is considered  to be 
of Lorentzian 
signature $(-,+,+,+)$.

\subsection{General very special relativity on cosmology}

A particularly interesting Finslerian cosmological model is elaborated in the framework 
of the so-called general very special relativity on cosmology
\cite{kour-stath-st 2010}. The metric function takes the 
form
\begin{equation}\label{vsgr metric}
F(x,y) = \big(g_{\mu\nu}(x)y^\mu y^\nu\big)^{(1-b)/2}\big(n_\kappa y^\kappa\big)^b,
\end{equation}
where $g_{\mu\nu}(x)$ is the  ordinary  FRW   metric (\ref{FRWdef}).
Expression \eqref{vsgr metric} is a direct cosmological 
generalization of the general very special relativity description, where the 
line-element is
\begin{equation}
ds = \big(\eta_{\mu\nu}dx^\mu dx^\nu\big)^{(1-b)/2}\big(n_\kappa dx^\kappa\big)^b,
\end{equation}
with $[\eta_{\mu\nu}] = \mathrm{diag}\big(-1,1,1,1\big)$, which is invariant under 
transformations generated by the deformation $DISIM_b(2)$ of the Lorentz subgroup 
$ISIM(2)$ \cite{Cohen:2006ky,Gibbons:2007iu}. The one-form $n_\kappa$ is 
called ``spurionic field''. We mention that the parameter $b$ quantifies the 
deviation from Riemannian geometry, i.e. the Lorentz 
violation in the gravitational sector. Parametrized post-Newtonian (PPN) analysis  
\cite{Will:2005va}  and 
use of solar system data provides the 
  most stringent constraints on it, and  thus
Gravity Probe B put an upper bound at $10^{-7}$ \cite{Bailey:2013oda}.
 
The Riemannian osculating approach is followed, namely $g_{\mu\nu}(x) = 
f_{\mu\nu}\big(x,y(x)\big)$, where $y(x)$ is the tangent vector to the cosmological 
fluid's (matter fluid) flow lines. As usual the matter fluid is described by the 
energy-momentum tensor of the perfect fluid:
\begin{equation}
\label{matterenergymomentum}
T_{\mu\nu} = P_mg_{\mu\nu} + (\rho_m+P_m)y_\mu y_\nu,
\end{equation}
where $\rho_m$ is the energy density and $P_m$ the pressure.
  The field equations for this construction are then:
\begin{equation}
L_{\mu\nu}-\frac{1}{2}Lg_{\mu\nu}=-8\pi G T_{\mu\nu}\label{feqs},
\end{equation}
where  $L_{\mu\nu}$ is the Ricci tensor for 
the metric $g_{\mu\nu}(x)$ and $L = g^{\mu\nu}L_{\mu\nu}$.

Applying the above geometrical construction in a cosmological framework we consider
the spurionic field   to be parallel to the 
velocity of the comoving observer, namely
\begin{equation}
n^\kappa = \big(n(t),0,0,0\big).
\end{equation}
As a simple model, in \cite{kour-stath-st 2010} the following approximations we imposed
\begin{eqnarray}
&& n(t) \approx At+B \nonumber\\
&& A \rightarrow 0\nonumber\\
&& B \rightarrow 0,
\label{GVSRappr}
\end{eqnarray}
since $n(t)$, parametrized by $A$,$B$, needs to be suitably small in order to be 
consistent with the observational small bound on $b$.
For these choices,   the Ricci tensor 
components for the 
metric function \eqref{vsgr metric} are calculated as \cite{kour-stath-st 2010}
\begin{eqnarray}
&&\!\!\!\!\!\!\!
L_{00}=  3\,\dfrac { \ddot{a }  }{a }+3\dfrac{Ab}{B}\dfrac{\dot{a}}{a} +O \left( 
{A}^{2} 
\right)  \nonumber\\
&&\!\!\!\!\!\!\!
L_{11}=  -\dfrac{ a\dot{a}+2\dot{a}^2+2k    
}{1-kr^2}+\dfrac{5A}{B}b\dfrac{a\dot{a} 
}{1-kr^2}+O(A^2)  \nonumber\\
&&\!\!\!\!\!\!\!
L_{22}=  -r^2( a\ddot{a}+2\dot{a}^2+2k     )- \dfrac{5A}{B}br^2a\ddot{a}+O(A^2) 
\nonumber\\
&&\!\!\!\!\!\!\!
L_{33}=  -r^2(a\ddot{a}+2\dot{a}^2+2k)\sin^2\theta-\dfrac{5A}{B}br^2 
a\ddot{a}\sin^2\theta\nonumber\\
&& \ \ \ \ \     \ +O(A^2).
 \label{riccit}
\end{eqnarray}
Therefore, using the above, we result to the following generalization of the Friedmann 
equations:
\begin{eqnarray}
&&
\!\!\!\!\!\!\!\!\!\!\!
H^2\!+\!\frac{k}{a^2}\!+\!2\frac{A}{B}b H =
\frac{8\pi G}{3}\! \left[ \rho_m \!-\!2\frac{A}{B}b P_m\!\left( t\! +\!\frac{B}{A}\ln B 
\right) 
\right] 
\label{Friedrad}\\
 &&\!\!\!\!\!\!\!\!\!\!\!
 \dot{H}+H^2 + \frac{Ab}{B}H = - \frac{4\pi 
G}{3}\big[(\rho_m+3P_m) \nonumber\\
&& \ \ \ \ \ \ \ \ \ \ \ \ \ \ \  \ \ \ \ \,  \ \ \ \ \ \ \  
+ 
4\ln(At+B)b(\rho_m+P_m)\big].
\end{eqnarray}
Unfortunately, as one can see, the above Friedmann equations do not accept a bounce 
solution. One could still try to construct a model with a different approximation than 
(\ref{GVSRappr}) of  \cite{kour-stath-st 2010}, however such a detailed investigation of 
a new construction lies beyond the scope of the present work. Hence, in the following 
subsection we examine the case of another Finslerian 
construction, where bounce realization is possible.

\subsection{Bounce in Finsler-Randers Space}

Let us now consider a different Finslerian construction, namely the Finsler-Randers (FR) 
space \cite{Stavrinos:2015,Stavrinos:2006rf}.  In this space a Lagrangian metric 
function is given by 
\begin{equation}
    F(x,y) = \upalpha(x,y) + u_\mu y^\mu, \, \| u_\mu \| \ll 1,
\end{equation}
 where $\upalpha(x,y)=\sqrt{g_{\kappa\lambda}(x)y^\kappa y^\lambda}$ and 
$g_{\kappa\lambda}(x)$ is the FRW metric \eqref{FRWdef}, with $\kappa,\lambda,\mu \in 
\{0,1,2,3\}$.
 
In this cosmological model an important role is played by the variation of anisotropy 
$Z_{t}$. In the case of the FRW geometry (\ref{FRWdef}) the modified Friedmann 
equations of the generalized form of the FR-type cosmology have been studied in 
\cite{Stavrinos:2006rf}, and are written as
\begin{eqnarray}\label{fr friedmann 1}
&&H^{2}=\frac{8 \pi G}{3} \rho_m-HZ_{t}   - 
\frac{k}{a^{2}},
\\
\label{fr friedmann 2}
&&\dot{H} = -4 \pi G
\left( \rho_m + P_m \right)+ \frac{1}{4} H Z_{t}+\frac{k}{a^{2}}.
\end{eqnarray}
In these expressions, we have 
defined the variation of anisotropy  $Z_{t}$ as $Z_{t}=\dot{u}_{0}$, namely as the 
derivative of the time component of the unit vector 
$\hat{u}_{a}$  \cite{Stavrinos:2006rf}.   This variation affects the form of 
geometry as   can be seen from relations \eqref{fr friedmann 1},\eqref{fr friedmann 2},  
and at the limit   $Z_t\rightarrow 0 $ we recover the ordinary Friedmann equations of 
general relativity. Finally, we have considered the matter sector to 
correspond to 
a perfect fluid with energy density and pressure $\rho_m$ and $P_m$ respectively.

Observing the form of two Friedmann equation (\ref{fr friedmann 1}),(\ref{fr 
friedmann 2})  we can define the effective energy density and pressure of geometrical 
origin as
\begin{eqnarray}
\label{rhoFR}
&&\rho_{FR}\equiv-\frac{3}{8\pi G} H Z_t
\\
\label{PFR}
&&P_{FR}\equiv\frac{5}{16\pi G} H Z_t.
\end{eqnarray}
Therefore,   the total energy density and pressure respectively become 
$\rho_{tot}=\rho_m+\rho_{FR}$ and $P_{tot}=P_m+P_{FR}$, and the Friedmann equations take 
the usual form of (\ref{FR1}),(\ref{FR2}). Hence, we can now easily examine what are the 
conditions in order to fulfill the bounce requirements (\ref{rhototcond1}) and 
(\ref{NULLene}).
 
Firstly, from   (\ref{rhototcond1})  we deduce that for the flat universe exactly at the 
bounce point $\rho_m$ must be zero ($\rho_{FR}$ becomes also zero exactly at the bounce 
point, since $H=0$). This is a usual assumption in many bouncing models and it is 
expected 
to be fulfilled in the early universe. Taking this into account, we moreover see that 
condition (\ref{NULLene}) implies that around the bouncing point $\rho_{FR}+P_{FR}<0$ and 
thus that $HZ_t>0$. Hence, we deduce that the above requirements can be fulfilled  if we 
suitably choose the   variation of anisotropy $Z_{t}$.

In order to provide a specific example 
we focus on a flat 
FRW geometry ($k=0$) and   we consider a bouncing scale factor of the form  
\begin{equation}
a(t)=a_b(1+Bt^2)^{1/3},
\label{bouncesf1}
\end{equation}
where $a_b$ is the scale factor value at the bounce, while 
$B$ is a positive parameter which determines how fast the
bounce takes place. In this case time varies between $-\infty $ and $+\infty $, 
with $t=0$ the bouncing point, and where away from the bounce one obtains the usual 
expansion behavior. Moreover, we consider that the matter sector  is absent in the early 
universe.
Inserting these into  (\ref{fr friedmann 1}) we immediately 
find that 
\begin{equation}\label{bounceFR2}
Z_{t}=- \frac{ 
 2   B t}{3    (1+ Bt^2)}.
\end{equation}
Hence, it is this $Z_t$, that comes from the Finslerian modification of the geometry, 
that generates the bouncing scale factor (\ref{bouncesf1}). Moreover, we remark that
the variation of anisotropy $Z_t$ actually determines the physically important quantity $B$  in 
(\ref{bouncesf1}).

\section{Finsler-like gravity on a tangent bundle}
\label{Finslertangentbundle}

In this Section we are interested in examining whether a bounce can be realized  
from Finsler-like gravity on a tangent bundle.  Generally, we will use the term 
Finsler-like for any metric theory in which the various structures may depend on a 
set of \emph{internal} variables ($y, \phi,$ etc) apart from the position or \emph{external} ones which we denote as $x^\mu$ through this work. Finsler-like extensions of general 
relativity on the tangent bundle have been presented in the bibliography \cite{bucataru-miron 2007,Vacaru:2005ht,miron-anastasiei 2012,pfeifer-wohlfarth 2012,Caianiello:1989wm} and bouncing cosmological scenarios have been studied on them \cite{Triantafyllopoulos:2018bli,Caianiello:1991,Gasperini:1991pp}. In the following we focus our 
interest on a tangent bundle $TM$ equipped with a Finslerian Sasaki-type metric:
\begin{equation}
\Gd = g_{\mu\nu}(x,y)\,\mathrm{d}x^\mu \otimes \mathrm{d}x^\nu + 
v_{\alpha\beta}(x,y)\,\delta y^\alpha \otimes \delta y^\beta \label{bundle metric},
\end{equation}
where $x^\mu$ are the coordinates on the base manifold, 
with $\kappa,\lambda,\mu,\nu,\ldots = 
0,1,2,3$, and $y^\alpha$ are the fiber coordinates, with $\alpha,\beta,\ldots,\theta = 
0,1,2,3$.  On the total space $TTM$ of $TM$, the adapted basis is 
$\{\delta_{\mu},\pdot\alpha\}$ and its dual is given by $\{\de x^\mu,\delta y^\alpha\}$, 
and the 
following definitions hold:
\begin{eqnarray}
&&
\delta_\mu = \dfrac{\delta}{\delta x^\mu} = \pder{}{x^\mu} - 
N^\alpha_\mu(x,y)\pder{}{y^\alpha} \label{delta x}\nonumber\\
&&\dot \partial_\alpha = \pder{}{y^\alpha}
\nonumber\\
&&\delta y^\alpha = \mathrm{d}y^\alpha + N^\alpha_\nu\mathrm{d}x^\nu \label{delta y},
\end{eqnarray}
where 
$N^\alpha_\mu(x,y)$ are the coefficients of a nonlinear connection on $TM$. This 
connection is defined by a splitting of the total space $TTM$ of $TM$ into an 
h-subspace $HTM$ spanned 
by $\{\delta_\mu\}$ and a v-subspace $VTM$ spanned by $\{\pdot \alpha\}$ 
\cite{Vacaru:2005ht}.
The tangent space of $TM$ is thus a Whitney sum of the h-subspace and v-subspace, namely
\begin{equation}
    TTM = HTM \oplus VTM.
\end{equation}

One can now introduce the $d-$connection $ 
\mathcal D $ as a covariant linear differentiation rule that preserves h-space and 
v-space:
\begin{align}
\mathcal D_{\delta_\kappa}\delta_\nu = L^\mu_{\nu\kappa}(x,y)\delta_\mu \qquad \mathcal 
D_{\pdot{\gamma}}\delta_\nu = C^\mu_{\nu\gamma}(x,y)\delta_\mu  \lin
\mathcal D_{\delta_\kappa}\pdot{\beta} = L^\alpha_{\beta\kappa}(x,y)\pdot{\alpha} \qquad 
\mathcal D_{\pdot{\gamma}}\pdot{\beta} = C^\alpha_{\beta\gamma}(x,y)\pdot{\alpha}.
\end{align}
A canonical $d-$connection is  a linear connection that is compatible with the metric 
\eqref{bundle 
metric}  and it preserves under parallel translation the horizontal and vertical 
subspaces $HTM$ and $VTM$ \cite{Vacaru:2005ht}. It can be uniquely defined if one demands 
that it only depends on $g_{\mu\nu}, 
v_{\alpha\beta}$ and $N^\alpha_\mu$, and moreover that the connection coefficients $ 
L^\mu_{\nu\kappa} $ and $ C^\alpha_{\beta\gamma} $ are symmetric on the lower indices. In 
this case, its coefficients turn out to be \cite{miron-watanabe-ikeda 1987}:
\begin{eqnarray}
&&
\!\!\!\!
L^\mu_{\nu\kappa}  = \frac{1}{2}g^{\mu\rho}\left(\delta_kg_{\rho\nu} + \delta_\nu 
g_{\rho\kappa} - \delta_\rho g_{\nu\kappa}\right) \label{metric d-connection 1}  
\nonumber\\
&&
\!\!\!\!
L^\alpha_{\beta\kappa}  = \dot{\partial}_\beta N^\alpha_\kappa + 
\frac{1}{2}v^{\alpha\gamma}\left(\delta_\kappa v_{\beta\gamma} - 
v_{\delta\gamma}\,\dot{\partial}_\beta N^\delta_\kappa - 
v_{\beta\delta}\,\dot{\partial}_\gamma N^\delta_\kappa\right) \label{metric d-connection 
2}  
\nonumber\\
&&
\!\!\!\!
C^\mu_{\nu\gamma}  = \frac{1}{2}g^{\mu\rho}\dot{\partial}_\gamma g_{\rho\nu} 
\label{metric d-connection 3} 
\nonumber\\
&&
\!\!\!\!
C^\alpha_{\beta\gamma}  = \frac{1}{2}v^{\alpha\delta}\left(\dot{\partial}_\gamma 
h_{\delta\beta} + \dot{\partial}_\beta h_{\delta\gamma} - \dot{\partial}_\delta 
v_{\beta\gamma}\right). \label{metric d-connection 4}
\end{eqnarray}
Now, the curvature of the nonlinear connection is defined as
\begin{equation}\label{Omega}
\Omega^\alpha_{\nu\kappa} = \dder{N^\alpha_\nu}{x^\kappa} - \dder{N^\alpha_\kappa}{x^\nu},
\end{equation}
and the space at hand is equipped with various Ricci curvature tensors such as:
\begin{align}
\overline R_{\mu\nu} & = \delta_\kappa L^\kappa_{\mu\nu} - \delta_\nu 
L^\kappa_{\mu\kappa} 
+ L^\rho_{\mu\nu}L^\kappa_{\rho\kappa} - L^\rho_{\mu\kappa}L^\kappa_{\rho\nu} \lin
S_{\alpha\beta} & = \pdot{\gamma}C^\gamma_{\alpha\beta} - 
\pdot{\beta}C^\gamma_{\alpha\gamma} + C^\epsilon_{\alpha\beta}C^\gamma_{\epsilon\gamma} - 
C^\epsilon_{\alpha\gamma}C^\gamma_{\epsilon\beta}. 
\label{d-ricci 4}
\end{align}
Hence, the generalized Ricci scalar curvature reads as
\begin{equation}
\R = g^{\mu\nu}\overline R_{\mu\nu} + v^{\alpha\beta}S_{\alpha\beta} \equiv \overline R+S.
\label{bundle ricci curvature}
\end{equation}

One can now write a Hilbert-like action, namely \cite{bucataru-miron 
2007,Vacaru:2005ht,miron-anastasiei 
2012,pfeifer-wohlfarth 2012}
\begin{eqnarray}\label{bundle action}
&&\!\!
\!\!\!\!\!\!\!\!
\mathcal S_{TM} = \dfrac{1}{16 \pi G}\mathcal S_{H} + \mathcal S_{M}\nonumber\\
&& \!\! \!\!\!\!\! \!\! \!\equiv  \dfrac{1}{16 \pi G}\!\int\! d^8\mathcal 
U\sqrt{\det\mathcal G}\,\mathcal{L}_{H}
+ \!\int \!d^8 \mathcal 
U\sqrt{\det\mathcal 
G}\,\mathcal{L}_{M}  ,
\end{eqnarray}
with
\begin{equation}
d^8 \mathcal U  = \de x^0\wedge\de x^1\wedge\de x^2\wedge\de x^3\wedge    \de 
y^0\wedge\de 
y^1\wedge\de y^2\wedge\de y^3 ,
\end{equation}
where the gravitational part of the action $\mathcal S_{H}$ is constructed by the 
gravitational 
Lagrangian 
\begin{equation}\label{bundle geometric action}
\mathcal{L}_{H}= \R =   (\overline R+S) ,
\end{equation}
and the matter action $\mathcal S_{M}$ by the matter Lagrangian $\mathcal L_M$.

Extremization of the total action $\mathcal S_{TM}$ with respect to the metric components 
$g_{\mu\nu}$ and 
$v_{\alpha\beta}$  leads to the following field equations 
\cite{Triantafyllopoulos:2018bli}:
\begin{gather}
\overline R_{(\mu\nu)} - \frac{1}{2}(\overline R + S) g_{\mu\nu} = 8\pi G T_{\mu\nu} 
\label{h field equations} \lin
S_{\alpha\beta} - \frac{1}{2}(\overline R + S) v_{\alpha\beta} = 8\pi G Y_{\alpha\beta},
\label{v field equations}
\end{gather}
where we have defined
$
T_{\mu\nu} = 
-\frac{2}{\sqrt{\det\mathcal{G}}}\frac{\delta{\left(\mathcal{L}_M\sqrt{\det\mathcal G 
}\right)}}{\delta g^{\mu\nu}} $ and $
Y_{\alpha\beta} = 
-\frac{2}{\sqrt{\det\mathcal{G}}}\frac{\delta{\left(\mathcal{L}_M\sqrt{\det\mathcal G 
}\right)}}{\delta v^{\alpha\beta}}$. Applying these field equations in the FRW metric 
(\ref{FRWdef}), focusing on the flat case, and assuming the usual matter perfect fluid 
(\ref{matterenergymomentum}), one obtains the following modified Friedmann equations 
\cite{Triantafyllopoulos:2018bli}: 
\begin{eqnarray}
&&H^2 = \frac{8\pi G}{3}\rho_m - \frac{1}{6}S 
\label{Fr1tangbundle} \\
&&
\dot{H}+H^2 = -\frac{4\pi G}{3} \left(\rho_m+3P_m\right) - \frac{1}{6}S 
\label{Fr2tangbundle},
\end{eqnarray}
where due to the imposed symmetries all quantities depend only on time.
 
From the form of the two Friedmann equations (\ref{Fr1tangbundle}),(\ref{Fr2tangbundle})  
we 
can see that we obtain extra contributions that reflect the Finsler-like structure of the
tangent bundle. In particular, these induce an effective energy density and pressure of 
geometrical origin as
\begin{eqnarray}
\label{rhotangbundle}
&&\rho_{S}\equiv-\frac{1}{16\pi G} S
\\
\label{Ptangbundle}
&&P_{S}\equiv\frac{1}{16\pi G} S.
\end{eqnarray}
Hence,   the total energy density and pressure respectively become 
$\rho_{tot}=\rho_m+\rho_{S}$ and $P_{tot}=P_m+P_{S}$, and the Friedmann equations acquire
the usual form of (\ref{FR1}),(\ref{FR2}). Thus, we can     examine what are the 
conditions in order to fulfill the bounce requirements (\ref{rhototcond1}) and 
(\ref{NULLene}). Concerning   (\ref{rhototcond1})  we deduce that for the flat universe 
exactly at the 
bounce point we must have $S=16\pi G\rho_m$, while (\ref{NULLene}) requires 
$\rho_{m}+P_{m}<0$ (since according to (\ref{rhotangbundle}),(\ref{Ptangbundle})  
$P_{S}+\rho_S=0$). Therefore, we conclude that in the case of a flat universe and for 
standard matter a bounce cannot be obtained in the scenario at hand.

  Nevertheless, a 
bounce could still be possible with the addition of extra fields, e.g. 
\cite{Triantafyllopoulos:2018bli}, but still one has to be careful with the constraints 
imposed to $S$ via \eqref{v field equations}. For example, if we consider the trivial case 
where $Y_{\alpha\beta} = 0$ then the trace of \eqref{v field equations} gives
\begin{equation}\label{Strivial}
    S=-2\overline R.
\end{equation}
We assume that the extra field can be modeled to a perfect fluid as in 
\eqref{matterenergymomentum}, with energy density and pressure $\rho_{eff}$ and $P_{eff}$ 
respectively, and thus the Friedmann equation \eqref{Fr1tangbundle} takes the form
\begin{equation}\label{fr extra field}
    H^2 = \frac{8\pi G}{3}(\rho_m + \rho_{eff}) - \frac{1}{6}S. 
\end{equation}
Substituting \eqref{Strivial} to \eqref{fr extra field}\footnote{In our case $\overline 
R$ reduces to the ordinary flat FRW Ricci scalar curvature of general relativity, due to 
the fact that the metric components $g_{\mu\nu}(x)$ do not depend on the $y$ 
variables, as was shown in \cite{Triantafyllopoulos:2018bli}.} gives $3H^2 +2\dot{H}+ 
8\pi G (\rho_m + \rho_{eff})/3 = 0$. This relation implies that in order for an extra 
field with trivial $Y_{\alpha\beta}$ to induce a bounce solution for our spatially flat 
metric it would need to have $\rho_{eff}<0$, which is undesirable from a physical point of 
view.

\section{Bounce from scalar-tensor theory on the fiber bundle}
\label{Finslerscalartensor}

In this section we investigate the bounce generation in theories which include 
scalar-tensor sectors on the fiber bundle. These constructions are very general, with 
very rich structure and behavior, which reveals the significant capabilities of 
Finsler-like 
geometry. We first present the basics of this 
construction and then we proceed to the  investigation of two explicit scenarios.

\subsection{The model}

We consider a fibered space over a pseudo-Riemannian spacetime manifold $M$ of the form 
$M 
\times 
\{\phi^{(1)}\} \times \{\phi^{(2)}\}$, where $ \phi^{(1)}, \phi^{(2)} $ stand for the 
fiber coordinates. Under coordinate transformations on the base manifold, fiber 
coordinates behave like scalars. Moreover, the space is equipped with a nonlinear 
connection with 
coefficients $N^{(\alpha)}_\mu(x^\nu,\phi^{(\beta)})$, where $\mu,\nu$ take the values 
from 
$0$ to $3$ and $\alpha,\beta$ take the values $1$ and $2$ \cite{stavrinos-ikeda 1999}. 
Its 
adapted bases for the 
tangent and cotangent spaces are $\{\delta_\mu = \partial_\mu - 
N^{(\beta)}_\mu\partial_{\phi^{(\beta)}}, \partial_{\phi^{(\alpha)}} \}$, where a 
summation 
is implied over the possible values of $\beta$, and $\{\de x^\mu, \delta\phi^{(\alpha)} = 
\de\phi^{(\alpha)} + N^{(\alpha)}_\mu\de x^\mu\}$ with a summation   implied over 
the 
possible values of $\mu$. The metric structure of the space is defined as 
\cite{stavrinos-ikeda 1999}
\begin{equation}
\mathbf G = g_{\mu\nu}(x)\,\de x^\mu\otimes\de x^\nu + 
v_{(\alpha)(\beta)}(x)\,\delta\phi^{(\alpha)}\otimes \delta\phi^{(\beta)}.
\end{equation}
The metric coefficients for the fiber coordinates are set as $v_{(0)(0)} = v_{(1)(1)} = 
\phi(x^\mu)$ and $v_{(0)(1)} = v_{(1)(0)} = 0 $. Note that the function $\phi$ is clearly 
a scalar under coordinate transformations.
The detailed investigation of the above construction has been performed in  
\cite{stavrinos-ikeda 1999}, where a metrical d-connection 
has been introduced and its curvature and torsion tensor coefficients have been 
calculated. Additionally, the Raychaudhuri equations for the model have been derived in 
\cite{Stavrinos:2016xyg}.

We can now write an action  as \cite{Stavrinos:2016xyg}
\begin{equation}
\mathcal S_G = \frac{1}{16\pi G}\int \sqrt{|\det\mathbf G|}\,\mathcal L_G dx^{(N)},
\end{equation}
where $\mathcal L_G$ is taken equal to the scalar curvature   of the 
d-connection, and 
$dx^{(N)} = d^4x\wedge\de\phi^{(1)}\wedge\de\phi^{(2)}$. In the special case of a 
holonomic basis, i.e. $[\delta_\mu,\delta_\nu]=0$, the scalar curvature of 
the 
d-connection is
\begin{equation}\label{holonomic R}
\mathcal R = R - \frac{2}{\phi}\square \phi + 
\frac{1}{4\phi^2}\partial^\mu\phi\partial_\mu\phi,
\end{equation}
where $R$ is the scalar curvature of Levi-Civita connection and $\square$ is the   
d'Alembert  operator with respect to it. On the other hand, in the 
general case one 
obtains 
the scalar curvature as
\begin{equation}\label{nonholonomic R}
\mathcal {\tilde R} = R - \frac{2}{\phi}\square \phi + 
\frac{1}{4\phi^2}\partial^\mu\phi\partial_\mu\phi + \frac{1}{\phi}\partial^\mu\phi\, 
\partial_{\phi^{(\alpha)}}N^{(\alpha)}_\mu.
\end{equation}

Additionally,     we can add the matter sector too, considering the total action
\begin{equation}
\mathcal S = \frac{1}{16\pi G} \int \sqrt{|\det\mathbf G|}\,\mathcal L_G dx^{(N)}+ \int 
\sqrt{|\det\mathbf G|}\,\mathcal L_M dx^{(N)}.
\label{totaction}
\end{equation}
Since for the determinants  $\det\mathbf G$ and $\det g$ we have the relation $\det\mathbf 
G=\phi^2 \det g $, the above total 
action can 
be re-written as
\begin{equation}
\mathcal S =  \frac{1}{16\pi G}\int \sqrt{|\det g|}\,\phi\mathcal L_G  dx^{(N)} + \int 
\sqrt{|\det g|}\,\phi\mathcal L_M dx^{(N)}.
\label{totaction2}
\end{equation}

In the following two subsections we study the bounce realization in the holonomic 
($\mathcal L_G=\mathcal R$) and nonholonomic ($\mathcal L_G=\mathcal {\tilde R}$) 
basis separately.

\subsection{Bounce in holonomic basis}

Let us consider the total action (\ref{totaction2}) in the case of holonomic basis, 
allowing 
also for a potential for the scalar field, namely 
\cite{Stavrinos:2016xyg}
\begin{eqnarray}
\mathcal S = && \!\!\!\!\!\!
\frac{1}{16 \pi G}\int\! \!\sqrt{|\det g|}\,\big[\phi\mathcal R - 
V(\phi)\big] dx^{(N)} \nonumber \\
&& \nonumber\\
&&\!\!\!\!\!\!\!
+ \int\!\! \sqrt{|\det g|}\,\phi\mathcal L_M dx^{(N)},
\end{eqnarray}
where $\mathcal R$ is the holonomic scalar curvature \eqref{holonomic R}.  
We mention here that the above action belongs to the Horndeski class, and hence the 
resulting equations of motion are guaranteed to have up to second order derivatives 
\cite{DeFelice:2011bh}. In particular, the field equations for the metric are 
extracted as
\begin{eqnarray}
&&\!\!\!\!\!\!\!\!\!\!\!\!\!\! 
E_{\mu\nu} = 8\pi G T_{\mu\nu} + \frac{1}{\phi}\big(\nabla_\mu\nabla_\nu\phi - 
g_{\mu\nu}\square\phi\big) 
\nonumber\\
&&\
+ \frac{1}{4\phi^2}\left[\frac{1}{2}g_{\mu\nu}(\nabla\phi)^2 - 
\nabla_\mu\phi\nabla_\nu\phi \right] - \frac{1}{2\phi}g_{\mu\nu}V,
\end{eqnarray}
where $E_{\mu\nu} = R_{\mu\nu} - \frac{1}{2}Rg_{\mu\nu}$ is the Einstein tensor, 
$T_{\mu\nu} = -\frac{2}{\sqrt{|g|}}\frac{\delta(\sqrt{|g|}\mathcal L_M)}{\delta 
g^{\mu\nu}}$ is  the energy-momentum tensor,  and $\nabla_\mu$ is the Levi-Civita 
covariant derivative, 
while the scalar field (extension of 
Klein-Gordon ) equation reads as
\begin{equation}\label{KG extension}
\square\phi = 2\phi\big( R-V'\big) + \frac{1}{2\phi}(\nabla\phi)^2 + 32\pi G\mathcal 
L_M\phi,
\end{equation}
with $V'=dV/d\phi$. Note the interesting fact that in the scenario at hand we obtain an 
effective interaction between the scalar field and the matter sector due to the 
transformation from $\mathbf G$-metric to   $g$-metric.

Applying the above equations in the  FRW metric  (\ref{FRWdef}), focusing on the flat 
case, and neglecting the matter sector, since we are interesting in the early-time bounce 
realization,  we obtain the following modified Friedmann equations: 
\begin{eqnarray}
\label{Fr1hol}
&&3H^2 =   -3H\frac{\dot\phi}{\phi} - \frac{\dot\phi^2}{8\phi^2}+\frac{1}{2\phi}V
\\
&&\dot{H}+H^2=  - \frac{1}{2\phi}\big(\ddot\phi + H\dot\phi\big) + 
\frac{\dot\phi^2}{12\phi^2} + \frac{V}{6\phi}
\label{Fr2hol}
\\
&&\ddot\phi + 3H\dot\phi = -12\phi\left(2H^2+\dot{H}\right)+ 
\frac{\dot\phi^2}{2\phi} +2\phi V'  ,
\end{eqnarray}
out of which two are independent.

We now proceed to show how it is possible to obtain a specific bounce in this 
construction. As we observe from the above equations, we may choose a specific 
scalar-field potential that can satisfy the general bounce conditions  
(\ref{rhototcond1}) 
and (\ref{NULLene}) and thus induce the bounce realization.  
We follow the procedure of  
\cite{Cai:2011tc,Cai:2012ag,Qiu:2013eoa,Cai:2010zma,Cai:2011bs} and we first start from 
the desired result, that is we
impose a known form of the scale factor $a(t)$ possessing a bouncing  
behavior. Thus, $H(t)$ is known too. Eliminating $V$ from (\ref{Fr1hol}),(\ref{Fr2hol})
gives the simple differential equation 
\begin{equation}
\label{Fr1holdiff}
 4\phi(t)\ddot{\phi}(t)-\dot{\phi(t)}[\dot{\phi}(t)+4H(t)\phi(t)]+8\dot{H}(t)\phi(t)^2=0,
\end{equation}
which can be solved to provide $\phi(t)$. Then this $\phi(t)$ can be inserted into  
(\ref{Fr1hol}) and provide $V(t)$ as
\begin{eqnarray}
\label{Fr1holVt}
V(t)=6 H(t)[\dot{\phi}(t)+\phi(t)H(t)]+\frac{\dot{\phi}(t)^2}{4\phi(t)}.
\end{eqnarray}
\begin{figure}[ht]
\fbox{\includegraphics[scale=0.415]{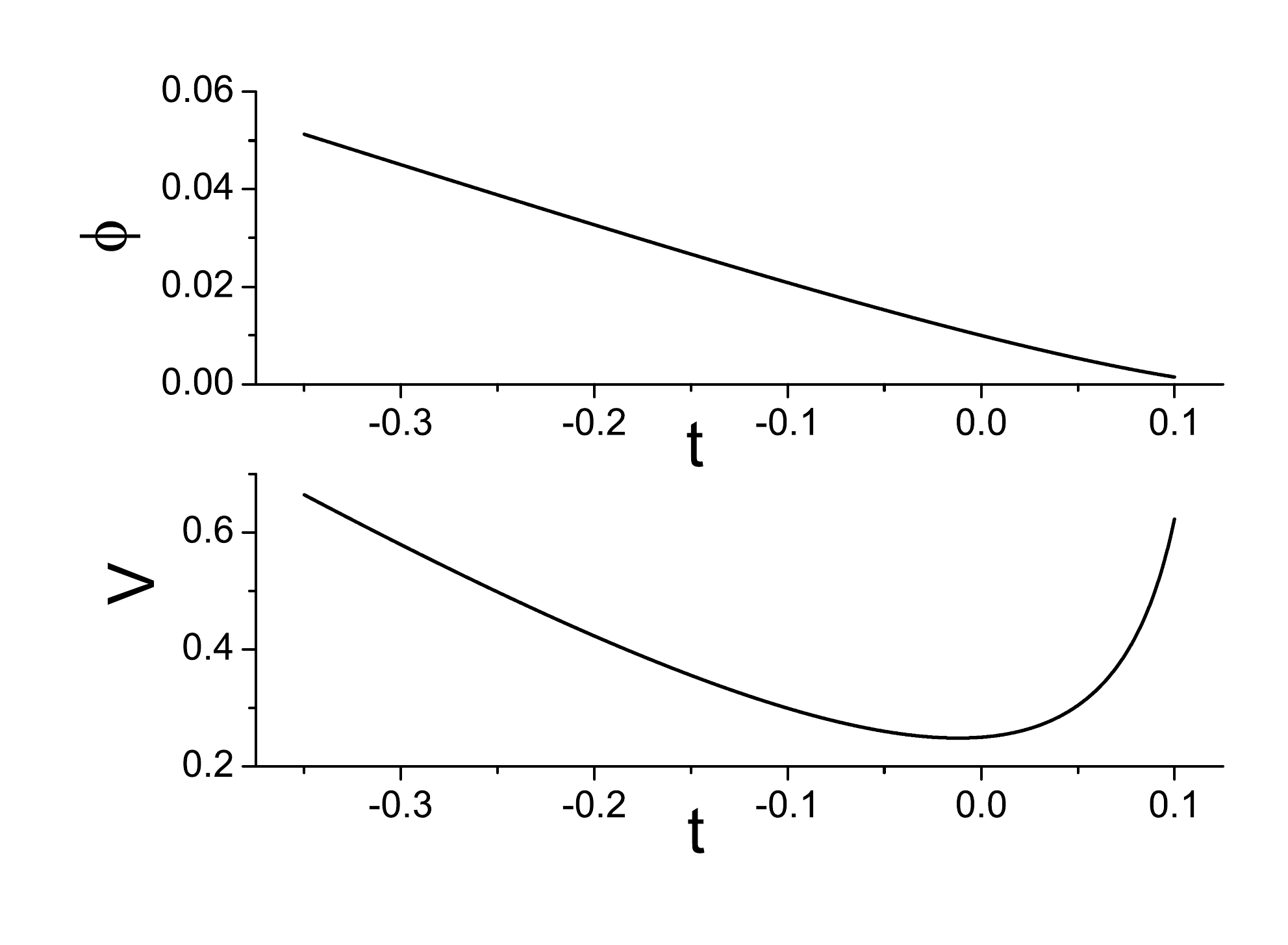}}
\caption{ 
{\it{ The solution for the scalar field $\phi(t)$ (upper graph) and of the potential 
$V(t)$ (lower graph), for the holonomic basis, under the imposed bouncing scale factor 
(\ref{bouncesf1}) with $B=1$, in units where $8\pi G=1$.
}} }
\label{Holon1}
\end{figure}
Finally, knowing both $\phi(t)$ and $V(t)$, eliminating time we can
extract the explicit form of the potential $V(\phi)$. Hence, it is this potential that generates the initially given desired bouncing   scale factor $a(t)$.

\begin{figure}[ht]
\fbox{\includegraphics[scale=0.43]{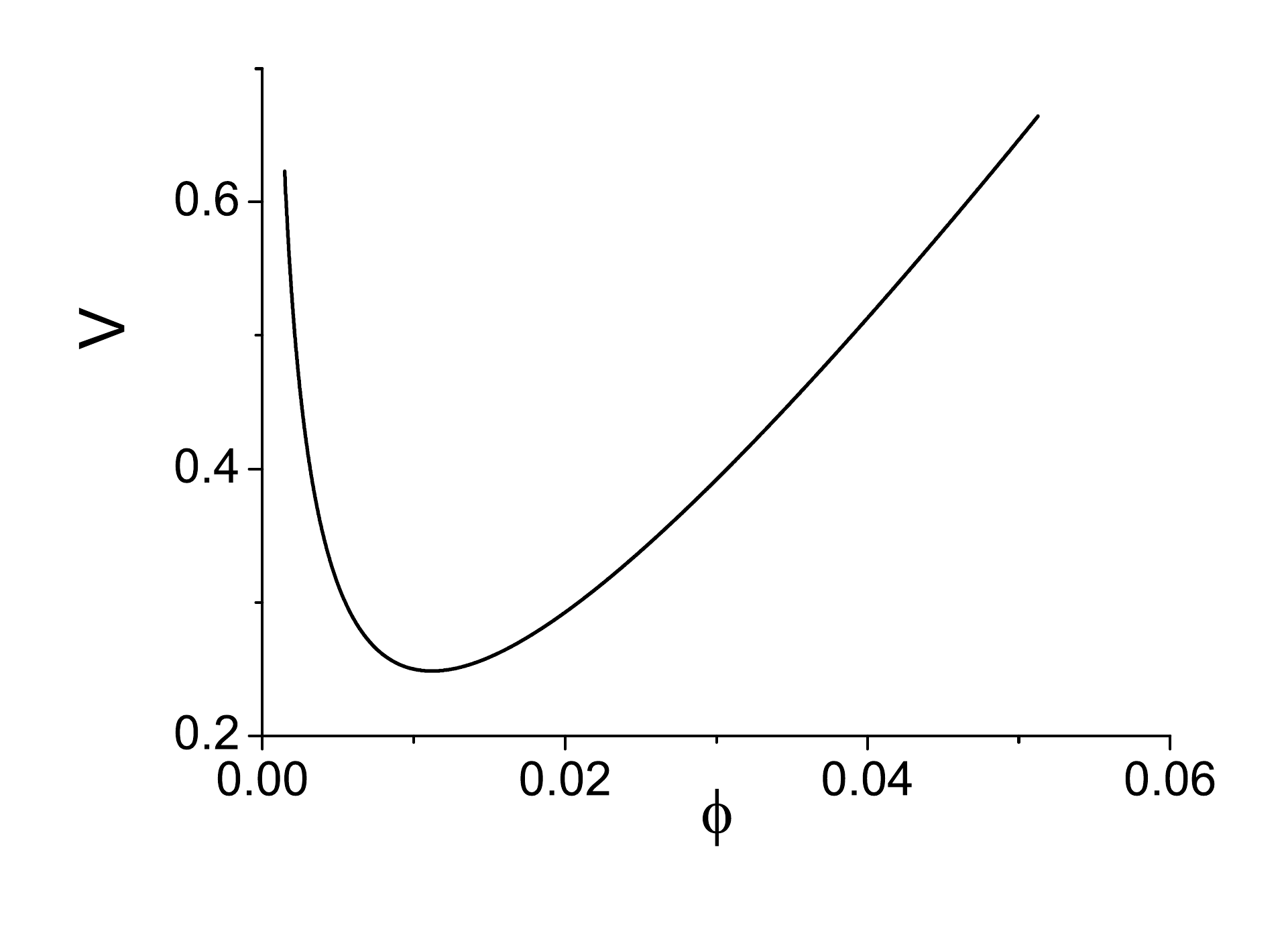}}
\caption{
{\it{ The reconstructed scalar potential $V(\phi)$ using Fig. \ref{Holon1},  under the 
imposed bouncing scale factor 
(\ref{bouncesf1}) with $B=1$, in units where $8\pi G=1$.
}} }
\label{Holon2}
\end{figure}

Let us provide an explicit example of the bounce realization. We start by inserting the 
desired bouncing scale factor (\ref{bouncesf1}) and we apply the above 
steps. Since analytical solutions cannot be obtained, we numerically solve 
(\ref{Fr1holdiff}) and find $\phi(t)$, and then we use (\ref{Fr1holVt}) to find $V(t)$.
These two functions are shown in Fig. \ref{Holon1}. Hence, from these  $\phi(t)$ and 
$V(t)$ we reconstruct the potential  $V(\phi)$, which is depicted in Fig. \ref{Holon2}.
Therefore, if this 
$V(\phi)$ is imposed as an input, one acquires the bounce realization, and in particular 
the bouncing scale factor (\ref{bouncesf1}).

\subsection{Bounce in nonholonomic basis}

We now proceed to the investigation of the nonholonomic case, namely 
we consider the total action (\ref{totaction2})  with $\mathcal L_G=\mathcal {\tilde R}$, 
i.e. 
 \begin{equation}
\mathcal S =  \frac{1}{16\pi G}\int \sqrt{|g|}\,\phi\mathcal{\tilde R} 
dx^{(N)} + \int \sqrt{|g|}\,\phi\mathcal L_M dx^{(N)},
\end{equation}
where $ \mathcal{\tilde R} $ is the nonholonomic scalar curvature \eqref{nonholonomic R}. 
This action leads to the following   equations of motion for the metric and the scalar 
field:
\begin{eqnarray}
&&
\!\!\!\!\!\!\!\!\!\!\!\!\!\!\!\!
E_{\mu\nu} = 8\pi G T_{\mu\nu} + \frac{1}{\phi}\big(\nabla_\mu\nabla_\nu\phi - 
g_{\mu\nu}\square\phi\big)
\nonumber\\
&&
+ \frac{1}{4\phi^2}\left[\frac{1}{2}g_{\mu\nu}(\nabla\phi)^2 - 
\nabla_\mu\phi\nabla_\nu\phi \right] \nonumber\\
&&- \left(\delta^\lambda_\mu\partial_\nu\phi - 
\frac{1}{2}g_{\mu\nu}\partial^\lambda\phi \right)N_\lambda \label{E unholonomic} \\
&&
\!\!\!\!\!\!\!\!\!\!\!\!\!\!\!\!
\square\phi = 2\phi R + \frac{1}{2\phi}(\nabla\phi)^2 + 32\pi G\mathcal L_M\phi - \phi 
D^\mu N_\mu ,
\label{phi unholonomic}
\end{eqnarray}
where $N_\mu \equiv \partial_{\phi^{(\alpha)}}N^{(\alpha)}_\mu$, and with $D_\mu 
N^\lambda = 
\delta_\mu N^\lambda + \Gamma^\lambda_{\kappa\mu}N^\kappa$   the d-covariant 
differentiation on the fiber bundle where $ \Gamma^\lambda_{\kappa\mu} $ are the 
Christoffel 
symbols. 
We mention that the last term in \eqref{phi unholonomic}, which reflects the internal 
structure of Finsler-like geometry, can be seen to act as an effective potential for 
the scalar field $\phi$. Since every other quantity in (\ref{E unholonomic}),(\ref{phi 
unholonomic}) depends on $x^\mu$ coordinates only, this should also be the case for  
$N_\lambda$   for consistency 
(equivalently $\partial_{\phi^{(\beta)}}\partial_{\phi^{(\alpha)}}N^{(\alpha)}_\mu = 0$ 
on shell).

Applying the above equations of motion in the  FRW metric  (\ref{FRWdef}), focusing on 
the flat 
case, and neglecting the matter sector, since we are interesting in the early-time bounce 
realization,   leads to the 
modified Friedmann equations
\begin{gather}
\label{FR1nonhol}
3H^2 = -3H\frac{\dot\phi}{\phi} - \frac{\dot\phi^2}{8\phi^2} - \frac{1}{2}\dot\phi N_0 
\lin
\dot{H}+H^2  =  - \frac{1}{2\phi}\big(\ddot\phi + H\dot\phi\big) + 
\frac{\dot\phi^2}{12\phi^2} + \frac{1}{3}\dot\phi N_0
\label{FR2nonhol}
\end{gather}
\begin{equation}
\ddot\phi + 3H\dot\phi = -12\phi\left(2H^2+\dot{H}\right) + 
\frac{\dot\phi^2}{2\phi}  + \phi\left(\dot N^0 + 3HN^0 \right),
\end{equation}
out of which two are independent,
where as we mentioned, due to symmetries, all quantities depend only on time. Thus, in 
the Friedmann equations we acquire a modification reflecting the nonholonomicity of the 
fiber bundle of the  underlying Finsler-like geometry.

 Let us now show how this construction may give rise to the bounce realization.
 From the form of the Friedmann equations (\ref{FR1nonhol}),(\ref{FR2nonhol}) we deduce 
that   we may choose a specific nonholonomic function $ N^0(t)$ that can satisfy the 
general bounce conditions  (\ref{rhototcond1}) 
and (\ref{NULLene}) and thus induce the bounce.  We first start from the desired 
result, that is we
impose as an input a scale factor form $a(t)$ that possesses bouncing  
behavior. Therefore, $H(t)$ is known too. Eliminating $ N^0$ from 
 (\ref{FR1nonhol}),(\ref{FR2nonhol})
gives the simple differential equation 
\begin{eqnarray}
\label{Fr1nonholdiff}
 \ddot{\phi}(t)+5 H(t)\dot{\phi(t)}  +2\phi(t)  
     [ \dot{H}(t)+3 H(t)^2]=0,
\end{eqnarray}
which can be solved to provide $\phi(t)$. Then this $\phi(t)$ can be substituted into  
(\ref{FR1nonhol}) and provide $N^0(t)$ as
\begin{eqnarray}
\label{Fr1nonholVt}
N_0(t)=-6 \left[\frac{H(t)}{\phi(t)}+\frac{\dot{\phi(t)}}{24 
\phi(t)^2}+\frac{H(t)^2}{\dot{\phi}(t)} \right].
\end{eqnarray} 
Hence, it is this  $N_0(t)$, induced by the nonlinear connection of the 
Finsler-like geometry, that 
generates the initially given desired bouncing   scale factor $a(t)$.

\begin{figure}[ht]
\fbox{\includegraphics[scale=0.45]{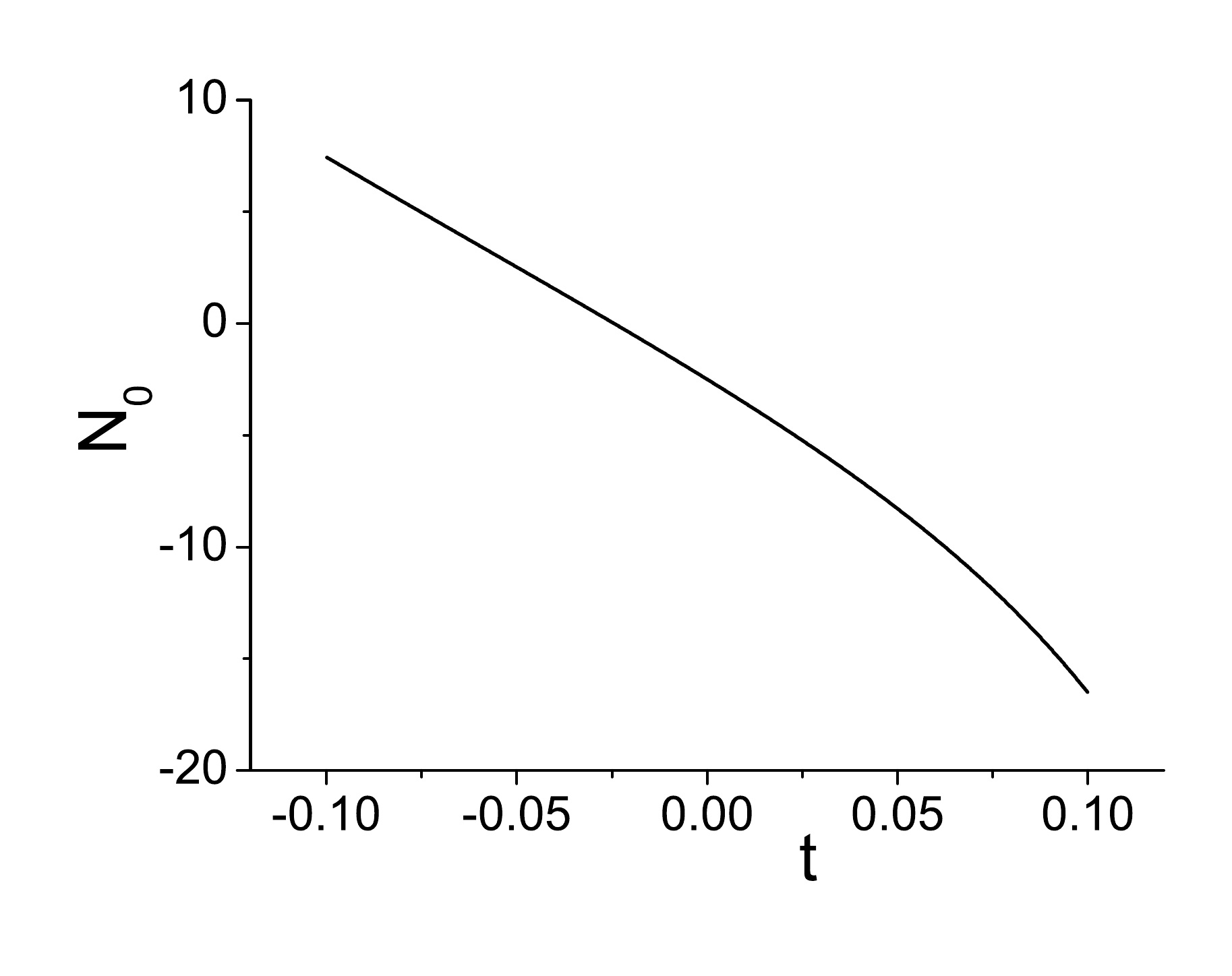}}
\caption{
{\it{ The reconstructed time-dependent part  $N_0(t)$ related to the nonlinear 
connection, for the nonholonomic basis,
 under the imposed bouncing scale factor 
(\ref{bouncesf1}) with $B=1$, in units where $8\pi G=1$.
}} }
\label{Nonholon}
\end{figure}

We close this subsection by providing an explicit example of the bounce realization. 
We use as input the bouncing scale factor (\ref{bouncesf1}) and we apply the above 
steps. We numerically solve 
(\ref{Fr1nonholdiff}) and find $\phi(t)$, and then we use (\ref{Fr1nonholVt}) to find 
 $N_0(t)$. In Fig. \ref{Nonholon} we depict the solution for $N_0(t)$. Hence, if this 
$N_0$ is imposed as an input, one obtains the bounce realization, and in particular the 
bouncing scale factor (\ref{bouncesf1}).

\section{Conclusions}
\label{Conclusions}

In this work we investigated the bounce realization in the framework of Finsler and 
Finsler-like 
gravity. Finsler and Finsler-like geometries are natural extensions of Riemannian 
one, where one allows that the physical quantities may directly depend on observer 
4-velocity. Hence, the gravitational theory based on  Finsler and Finsler-like 
gravity provides a gravitational modification, since it induces extra terms in the field 
equations. When applied in a cosmological framework, the richer intrinsic 
structure of Finsler and Finsler-like geometries is reflected in extra terms in the 
resulting modified 
Friedmann equations. Thus, these terms can lead to the bounce realizations.

In our analysis we considered various Finsler and Finsler-like constructions and we 
examined whether bouncing solutions can be obtained. As a first model we considered 
the so-called general very special relativity, which presents a slight Lorentz violation 
quantified by a single parameter and the ``spurionic'' one-form. As we showed, under the 
linear approximation this scenario cannot lead to a bounce. However, considering the 
Finsler-Randers space, in which the intrinsic Finslerian structure is reflected to the 
appearance of a new function in the Friedmann equations (the variation of anisotropy), we 
saw that the bounce conditions can be easily fulfilled and thus the bounce can be 
realized.

As a next construction we examined the Finsler-like gravity on the tangent bundle. 
Performing the analysis and considering the two involved curvature tensors, we extracted 
the Friedmann equations which contain a modification resulting from the tangent-bundle 
related $S$-curvature. Nevertheless, for simple models and standard matter, these extra 
terms cannot drive a bouncing solution since they cannot lead to the violation of the 
null energy condition.

As a last construction we considered theories which include scalar-tensor sectors on the 
fiber bundle. These theories present a very rich structure  revealing the capabilities of 
Finsler-like geometry. In particular, the nonlinear connection induces a new degree of 
freedom that behaves as a scalar under coordinate transformations. In a cosmological 
framework this scalar field appears in the Friedmann equations, and therefore its 
dynamics may trigger a bounce. In the case of holonomic basis we showed that the bounce 
can be easily obtained, and we provided the way of the reconstruction of the potential 
that gives rise to a  desired bouncing scale factor. Similarly, in the 
case of nonholonomic basis we saw that the bounce can be easily realized, and we 
presented the reconstruction procedure of the time-coefficient related 
to  the  nonlinear connection that 
induces the desired bounce.

In summary, we saw that  Finsler and Finsler-like geometries is a natural framework for the 
realization of 
bounce cosmology. Apart from the background evolution one should additionally investigate 
the various scenarios at the perturbation levels, since  the process of perturbations 
through the bounce phase is strongly related to the subsequent development of the large 
scale structure and hence to observations. Such a detailed perturbation analysis lies 
beyond the scope of the present work and it is left for a future investigation.

\begin{acknowledgments}
This research is co-financed by Greece and the European Union (European Social Fund- ESF) 
through the Operational Programme ``Human Resources Development, Education and Lifelong 
Learning'' in the context of the project ``Strengthening Human Resources Research 
Potential via Doctorate Research'' (MIS-5000432), implemented by the State Scholarships 
Foundation (IKY). This article is based upon work from COST (European Cooperation in 
Science and Technology) Action CA15117 ``Cosmology and Astrophysics
Network for Theoretical Advances and Training Actions'' (CANTATA).
\end{acknowledgments}

\end{document}